\documentclass[letterpaper, 10 pt, conference]{ieeeconf}  
\newif\ifdraft
 \draftfalse
\IEEEoverridecommandlockouts                              
\usepackage{graphicx}
\usepackage[
    backend=biber,
    style=ieee,
    sorting=none
]{biblatex}
\renewbibmacro*{bbx:savehash}{}

\bibliography{myBib}
\usepackage[hidelinks]{hyperref}
\usepackage{xcolor}
\usepackage[acronym]{glossaries}
\usepackage{amsmath, amsfonts, amssymb, mathtools, ntheorem}  
\usepackage{bm}
\usepackage[nameinlink,capitalise]{cleveref}
\usepackage{xspace}
\usepackage{lipsum}
\usepackage{bbm}
\usepackage{algorithm}
\usepackage[noend]{algpseudocode}
\usepackage{nicematrix,tikz}
\usetikzlibrary{fit}
\usepackage[normalem]{ulem}
\usepackage{subcaption}
\usepackage{titlesec}
\titlespacing\section{0pt}{2pt minus 1pt}{2pt minus 1pt}
\titlespacing\subsection{0pt}{2pt minus 1pt}{2pt minus 1pt}

\definecolor{cricolor}{HTML}{FF9999}
\definecolor{stephaniecolor}{HTML}{19e194}
\definecolor{resultscolor}{HTML}{c71585}
\definecolor{aroncolor}{HTML}{FFB612}

\newcommand{\cnote}[1]{{\ifdraft\color{cricolor} Cris: #1\fi}}

\newcommand{\anote}[1]{{\ifdraft\color{aroncolor} Áron: #1\fi}}
\newacronym{sbm}{SBM}{Stochastic BlockModel}
\newacronym{mca}{MCA}{Median Consensus Algorithm}
\newacronym{cmca}{C-MCA}{Community-Based Median Consensus Algorithm}
\newacronym{rac}{RAC}{Resilient Asymptotic Consensus}
\newacronym{dsn}{DSN}{Distributed Sensor networks}
\newacronym{das}{DAS}{Distirbuted Acoustic Sensing}
\newacronym{iot}{IoT}{Internet of Things}

\hypersetup{
    colorlinks,
    linkcolor={black},
    citecolor={teal!50!black},
    urlcolor={black}
}

\newcommand{\badval}{\bold{b}\xspace}

\newcommand{\neigh}[1]{\mathcal{N}_#1}

\newcommand{\phaseone}{\emph{phase1}\xspace}
\newcommand{\phasetwo}{\emph{phase2}\xspace}
\newcommand{\Phaseone}{Phase1\xspace}
\newcommand{\Phasetwo}{Phase2\xspace}

\newcommand{\median}{M}

\newcommand{\redundantvals}{\mathcal{R}}
\newcommand{\subs}{V}
\newcommand{\kedges}[1]{k#1}

\newcommand{\structuraledges}{\emph{structural edges}\xspace}
\newcommand{\oracledges}{\emph{oracle edges}\xspace}
\newcommand{\oracleprob}{p_O\xspace}
\newcommand{\fixedpoint}[1]{\mathcal{M}_{#1}}
\newcommand{\leftarrowtext}[1]{\xleftarrow{\text{#1}}}
\newcommand{\avg}[1]{\text{Avg}({#1})}
\newcommand{\ginduced}[1]{G[\subs#1]}
\newcommand{\oracle}{\emph{Oracle}\xspace}
\DeclareMathOperator*{\argmax}{\arg\!\max}

\newtheorem{theorem}{Theorem}

\newtheorem{definition}{Definition}
\newtheorem{proposition}{Proposition}
\newtheorem{remark}{Remark}
\newcommand{\qed}{\hfill\blacksquare}

\title{\LARGE \bf 
Community Consensus: Converging Locally despite Adversaries and Heterogeneous Connectivity}

\author{ \parbox{3 in}{\centering Huibert Kwakernaak*
        \thanks{*Use the $\backslash$thanks command to put information here}\\
        Faculty of Electrical Engineering, Mathematics and Computer Science\\
        University of Twente\\
        7500 AE Enschede, The Netherlands\\
        {\tt\small h.kwakernaak@autsubmit.com}}
        \hspace*{ 0.5 in}
        \parbox{3 in}{ \centering Pradeep Misra**
        \thanks{**The footnote marks may be inserted manually}\\
       Department of Electrical Engineering \\
        Wright State University\\
        Dayton, OH 45435, USA\\
        {\tt\small pmisra@cs.wright.edu}}
}

\author{Cristina Gava, \'{A}ron V\'{e}k\'{a}ssy, Matthew Cavorsi,  Stephanie Gil, Frederik Mallmann-Trenn
\thanks{Cristina Gava and Frederik Mallmann-Trenn are with the Department of Informatics, King’s College London, WC2R 2LS London, U.K. (e-mail:{\small frederik.mallmann-trenn@kcl.ac.uk}; {\small cristina.gava@kcl.ac.uk}).
\'{A}ron V\'{e}k\'{a}ssy, Matthew Cavorsi and Stephanie Gil are with the School of Engineering, Applied Sciences at Harvard University, Allston, MA 02134 USA (e-mail: {\small avekassy@g.harvard.edu}; {\small mcavorsi@g.harvard.edu}; {\small sgil@seas.harvard.edu}).\newline The authors gratefully acknowledge AFOSR grant \#FA9550-22-1-0223, NSF grant \#CNS-2147694 and EP/W005573/1 for partial support of this work.}
}

\begin{document}

\maketitle

\begin{abstract}

We introduce the concept of \emph{community consensus} in the presence of malicious agents using a well-known median-based consensus algorithm. We consider networks that have multiple well-connected regions that we term \emph{communities}, characterized by specific robustness and minimum degree properties. Prior work derives conditions on properties that are necessary and sufficient for achieving \emph{global} consensus in a network. This, however, requires the minimum degree of the network graph to be proportional to the number of malicious agents in the network, which is not very practical in large networks. In this work, we present a natural generalization of this previous result. We characterize cases where, although global consensus is not reached, some subsets of agents $\subs_i$ will still converge to the same values $\mathcal{M}_i$ among themselves. To reach this new type of consensus, we define more relaxed requirements in terms of the number of malicious agents in each community, and the number $\kedges$ of edges connecting an agent in a community to agents external to the community. 
\end{abstract}
Keywords: Consensus, Distributed Model, Malicious Agents, MCA, Communities

\section{Introduction} \label{sec:intro}
In this work we present a new form of local consensus in a network, 
that we call \emph{community consensus},
that can be achieved even in the presence of malicious agents. We focus on the case where legitimate agents in different ``communities'' of a network try to reach consensus within their respective subgraph. We characterize conditions on the graph topology such that community consensus can be reached despite the presence of malicious agents. Importantly, community consensus can still be attained in many networks where global consensus ~\cite{medianSundaram} is not possible. 

The need for multiple agents to agree on a value in a distributed manner is crucial for many real-life applications, from systems of multiple sensors to social dynamics. The topology and connectivity of the network, though, can impede that agents receive information from any region of the network, preventing common agreement. Conversely, there can be scenarios where the ultimate goal is not to have an entire network to agree on a unique value, but instead having different regions of the network to agree on different values -- which is the focal point of this paper. 
We consider a network where agents communicate in a distributed manner and they can query the values of all their neighbors. However,
they do not know whether other agents are malicious nor the structure of the network outside of their neighborhood. 

Many works in the literature focus on consensus problems in networks under a graph-theoretic and stochastic perspective, from average consensus \cite{Cao2005, cao2008reaching} to opinion dynamics \cite{LORENZ2005217, Degroot1974}. Even malicious agents are considered in some works (\cite{pasqualetti2011consensus, yemini2021characterizing, gil2018resilient, Wang2019}), but little attention has been posed on networks having heterogeneous connectivity and malicious agents.
Some of these works also focus on devising strategies to identify malicious agents or exclude them from the process of reaching consensus \cite{parlangeli2021detection, FMS22}.
Here, identifying malicious agents is not a primary focus, instead, we focus on characterizing conditions to reach community consensus with unknown malicious agents in the network. 

We build upon the model devised in \cite{medianSundaram}, which aims for all legitimate agents in the network to converge to the same value. In~\cite{medianSundaram}, the authors require a strong notion of robustness that may not be applicable to many practical use cases, where these stringent connectivity constraints are not fulfilled. We use the same median consensus algorithm, but pose a different goal, in that we aim for legitimate agents within a subgraph of the network to converge to a value that is within the convex hull of their initial values, i.e., \emph{community consensus}. Relaxing the goal also allows us to relax the connectivity constraints. However, reaching community consensus is non-trivial for two important reasons: 1) consensus still requires sufficient robustness within a community, but also 2) community consensus requires some degree of isolation between communities.\cnote{We did not define what we mean by \emph{isolation} yet here. Perhaps add something like "i.e. values that are outside the convex hull of initial values present in one specific community should not influence agents of that community."} \anote{Here it is just here in its regular meaning as a standalone word, to give a sense that communities cannot be connected too much to each other}\cnote{ok sounds good} Specifically, a characterization of connectivity within the community and across communities that depends on the number of malicious agents in each community is required. 

Towards the goal of characterizing conditions when community consensus can be achieved, our work addresses several challenges: for some community $i$, legitimate agents in a subset $\subs_i$ of the network need to be able to reach the same fixed point $\fixedpoint{\subs_i} \in \mathbb{R}$, and they need to be robust outliers, with respect to the values in $\subs_i$. For the edge case of the entire network being a single community, \cite{medianSundaram} already provides necessary and sufficient conditions for this to happen. However, with multiple communities connected to each other, our analysis has to take into account the effect these communities can have on each other. In this work, we characterize the parameters of a network $G$ such that community consensus is attainable. As we will show in \cref{sec:results}, this is done in terms of the number of edges $\kedges_i$ connecting agents in subset $\subs_i$ to other subsets, and in terms of the number of malicious nodes $f_i$ in $\subs_i$. The strength of our approach is that it models scenarios in which the network connectivity is not homogeneous. 

This paper is structured as follows. In \cref{sec:literature}, we  provide the current state of the art and explain how our work  places itself in it, also recalling, in \cref{sec:robustness}, important notions of network robustness that we make use of. Our framework is described in \cref{sec:problemformulation} and in \cref{sec:algorithm}, while in \cref{sec:results} we present our theoretical results and we prove them. Empirical results are presented in \cref{sec:simulations}, precisely showing the necessity for the constraints in our main theorem. We finish this work with describing some relevant use cases in \cref{sec:usecases} and drawing final conclusions in \cref{sec:conclusions}.

\section{Related Literature}\label{sec:literature}
Consensus problems on evolving networks have been thoroughly studied in many fields, from opinion dynamics (See for example \cite{Degroot1974, LORENZ2005217, lorenz2007repeated} and the references therein) to first order distributed optimization algorithms \cite{Jadbabaie2003, OSaber2004}, where linear matrix equations and Lyapunov functions are used. These works present the so-called \emph{average consensus} protocols, where agents update their value by averaging it with the value of their neighbors.
Works on average consensus have thoroughly developed under many points of view: From fixed and switching topologies, \cite{Cao2008, Ming2008, vicsek1995novel}, to a graphical approach  \cite{cao2008reaching}, and to a Markov Chain approach \cite{Cao2005, berenbrink2023distributed}, where worst-case convergence rates and graph-theoretic conditions have been examined.
Aside from the average, other aggregates prove useful and more robust to outliers.
Fundamental is the seminal work from Kempe \cite{Kempe2003}, which, with probability $1- \delta$, shows an $\varepsilon$-convergence in the \emph{push-sum} gossip-based protocol in $O(\log{n})$. In it, authors study the problems of computing several types of aggregates through gossip-based protocols. A follow up work is \cite{Kuhn2007}, where the authors focus on an asynchronous way to find the $\kedges^{th}$ smallest value in a network of $n$ agents. In this work, authors concentrate on the median, which is a valuable aggregate, because of its robustness to outliers. 
In the control theory community, some works explored different robust aggregates besides median and mean: some methods extend to tools such as Krum and multi-Krum \cite{blanchard2017machine}, geometric median \cite{chen2017distributed}, coordinate-wise median,
coordinate-wise trimmed mean \cite{yang2019byrdie}, 
Bulyan and multi-Bulyan \cite{guerraoui2018hidden}.
A fundamental aspect of our work is the presence of adversarial agents in the network. While the presence of wrong values in a network has been previously considered, the literature around consensus models with adversaries (i.e., the number of malicious or spoofed agents) is still relatively understudied.
Many recent works on adversarial agents pose the attention on leader-follower dynamics \cite{Wang2019} where the value of one agent is taken as reference for the other agents to follow. Our work applies to a leaderless dynamic, where agents possibly reach a common value but are unaware of the global status of the network (decentralized approach) and of the fixed point. The existing results on resilient cooperative control are, however, still conditioned on many communication network’s connectivity requirements.
In the case of \cite{angluin2008simple}, the authors show robustness of the network of size $n$ to $o(\sqrt{n})$ byzantine agents, while \cite{Doerr2011} looks at the case up to $\sqrt{n}$ agents can be corrupted in the network at any time. In \cite{Angluin2006} authors design what they call a ``Stabilizing Consensus'' able to tolerate $f<n$ crash faults and $f \; s.t. \; 3f <n$ byzantine faults in a network of $n$ agents.
These works leverage on asynchronous models, and there is no focus on the underlying network topology, but instead, interaction among agents is modeled as random matchings. Our work compares to the work of Zhang and Sundaram \cite{medianSundaram} and builds from that. The authors look at a synchronous, distributed, median consensus model where $f$ malicious agents can be present in a network of $n$ agents.
To show the validity of the model, authors devise a new notion of robustness, namely \emph{(r,s)-excess robustness} and pair it with the requirement that the minimum degree of the network be $d_{min} \ge 2f + 1$. Their work is a further development from previous work in \cite{wmsrsundaramc} and \cite{haotian2012}.
\section{Background: Robustness} \label{sec:robustness}
\subsection{The \texorpdfstring{\gls{mca}}{}}
The authors of \cite{medianSundaram} present a consensus protocol where agents in a network follow a synchronous, distributed consensus algorithm called \gls{mca}. We refer to it as \emph{median-based consensus protocol}. In the \gls{mca}, every legitimate agent $u$ in a graph $G$ holds a value. At every step $t$, this value is averaged with the median of the values from $u$'s neighbors. The algorithm iterates perpetually and the authors prove that the agents will eventually reach a type of consensus they call \emph{\gls{rac}} (defined in \cref{def:resilientconsensus}). Due to the presence of malicious agents, legitimate agents can only reach consensus under certain assumptions of robustness of the graph. To this end, \cite{medianSundaram} introduces \emph{$(r,s)$-excess robustness}, building from the notion of \emph{$r$-excess robustness} in the previous works. 
The notion of $r$-excess robustness conveys the idea that, for any partition of the network, at least one agent in the partition has $r$ more neighbors outside the partition than inside; $(r, s)$-excess robustness  asks that there always be at least $s$ of such agents.
We recall here the notions of robustness presented in \cite{medianSundaram}.
In our approach, we use the same notions and observe their application to sub-graphs induced
in the original graph.\footnote{We recall here that an \emph{induced sub-graph} on $G = (V, E)$ is a graph $\ginduced{^*} = (V^*, E^*)$ where $V^* \subseteq V$ and $E^*$ contains \emph{all} of the edges, from the original graph $G$, that connect the agents in $V^*$.}

%
\subsection{\texorpdfstring{$r$-excess robustness}{} and \texorpdfstring{$(r, s)$-excess robustness}{}}
The median is a robust statistic that is not influenced by outliers. However, this also means that legitimate values at the extremes of an ordered vector of values may never be selected as the median.
Consider the example of a graph $G$ where agents can be partitioned into two sets $\mathcal{A}$ and $\mathcal{B}$, such that, at time $t=0$, any agent $u \in \mathcal{A}$ holds a value $a$ and any agent $v \in \mathcal{B}$ holds $b \neq a$. Assuming agents share and update their values following the \gls{mca}, they will not reach convergence if every agent has more neighbors in its own subset than neighbors in the other subset -- no agent will change its opinion, regardless of the connectivity in other parts of the graph.
Stronger assumptions on connectivity are hence needed, as provided by the notion of $r$-excess robustness, from \cite{wmsrsundaramc} and \cite{medianSundaram}. Let $\mathcal{N}_u$ denote the neighborhood of  agent $u$.

\begin{definition}[$r$-excess reachable set]\label{def:rreachable}
    Given a graph $G = (V, E)$ and a nonempty subset $S$ of agents of $G$, we say $S$ is an $r$-excess reachable set if $\exists u \in S$ such that $|\mathcal{N}_u \setminus S| - |\mathcal{N}_u\cap S| \geq r$, where $r \in \mathbb{Z}_{\geq 0}$. When clear from the context, we will also say that agent $u$ (with regard to set $S$) is $r$-excess reachable. 
\end{definition}

\begin{definition}[$r$-excess robust graph]\label{them:rrobust}
A graph $G = (V, E)$ is $r$-excess robust, with $r \in \mathbb{Z}_{\geq 0}$, if for every pair of nonempty, disjoint subsets of $V$, at least one of the subsets is $r$-excess reachable.
\end{definition}
However, it  might happen that all the reachable agents are also malicious, resulting in the network being disconnected and preventing its legitimate agents from reaching consensus. To address this case, in \cite{medianSundaram}, robustness is expanded to $(r,s)$-excess robustness, with the idea to ensure that there always are more reachable agents than malicious agents. 

\begin{definition}[$(r, s)$-excess robustness]\label{thm:rsrobust}
Take $r \in \mathbb{Z}_{\geq 0}$ and $s \in \{1, . . . , n\}$. A graph $G = (V, E)$ is $(r,s)$-excess robust if
\begin{itemize}
    \item For every pair of nonempty, disjoint subsets $S_1, S_2 \subseteq V$
    \item Given the set $\mathcal{X}^r_{S_i}= \{u \in S_i : |\mathcal{N}_u \backslash S_i| - |\mathcal{N}_u \cap S_i| \geq r\}$ for $i \in \{1,2\}$
    \end{itemize}
At least one of the following holds: $|\mathcal{X}^r_{S_1}| + |\mathcal{X}^r_{S_2}| \geq s$; $|\mathcal{X}^r_{S_1}| = |S_1|$; $|\mathcal{X}^r_{S_2}| = |S_2|$.
\end{definition}
\section{Problem Formulation} \label{sec:problemformulation}
The previous definitions leverage on the property that all the agents in the graph need to be somewhat \emph{connected enough} in order for the \gls{mca} to work. In other words, this means assuring that any subset of agents $S\subseteq V$, as defined in \cref{def:rreachable}, is able to incorporate values from agents in $V \setminus S$ in the update step.
In this section, we apply the notions of consensus and robustness from \cite{medianSundaram} to more general networks, where robustness properties pertain to induced sub-graphs of the starting graph $G$, rather than its entirety.
We formalize the case of a graph with heterogeneous connectivity, where more connected agents belong to subsets that fulfill $(r, s)$-excess robustness criteria. In these subsets, legitimate agents can reach consensus and not only are they robust to malicious agents in the same subset, but to malicious agents in the whole $G$ as well.

\subsection{Graph Structure and Additional Notation}\label{sec:graph} We consider a graph $G = (V, E)$. Let $n=|V|$. In $G$, we denote by \emph{legitimate} the agents that follow the \gls{mca} and by \emph{malicious} all the others. We call $L$ the set of legitimate agents in $G$ and $F$ the set of malicious agents,  such that $F \cup L = V$ and $F \cap L = \emptyset$. We define a partition of $V$: $\{\subs_1, \subs_2, \dots, \subs_c\}$ into $c$ subsets, where $\subs_i$ indicates the generic $i^{th}$ subset.
\begin{figure}
    \centering
    \includegraphics[width=.6\columnwidth]{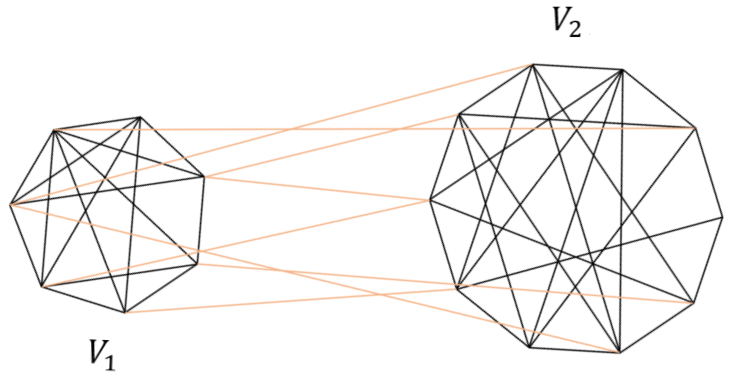}
    \caption{Example of graph where two subsets of agents $V_1$ and $V_2$ are visibly more connected within themselves. Lower connectivity between $\subs_1$ and $\subs_2$ is represented by the orange edges.}
    \label{fig:commExample}
\end{figure}
For any subset $\subs_i \subseteq V$ and an agent $u \in \subs_i$, we call $\mathcal{N}_u = \{v \ | \ (u, v) \in E\}$ and $\kedges_i = \max_u\{|\mathcal{N}_u \setminus \subs_i|\} \ge 0$.
We use the notation $\ginduced{_i}$ to indicate the sub-graph of $G$ induced by the subset of agents in $\subs_i$.
Further, an agent $u$ has degree $d_u$, $G$'s minimum degree is denoted by $d_{min} = \min_u\{d_u\}$, and the minimum degree of $\ginduced{_i}$ is denoted by $d_{min}^{\subs_i}$.

The process unfolds in discrete rounds, starting from $t=0$. We represent with $\xi_u(t) \in \mathbb{R}$ the value of an agent $u$ at time $t$, let it be legitimate or malicious. We assume that legitimate agents do not know which agents are malicious, nor which values they can take. However, malicious agents may know each other and, therefore, distinguish between legitimate and other malicious agents. 

We do not make assumptions on the malicious values, allowing  us to account for the fact that malicious agents can be able to coordinate their values in such way that their action is the most disruptive. This includes being able to potentially update their value in later stages. 
Recall that $\neigh{u}$ indicates the neighborhood of $u$, we call $\boldsymbol{\xi}^{(u)}(t) = \{\xi_v(t): v \in \neigh{u}\}$ the \emph{ordered} vector of values held by the neighbors of $u$ at time $t$. When there is no ambiguity, we will omit the time reference $t$. Finally, using the notation $\boldsymbol{\xi}^{(u)}_{[i]}$ to indicate the $i^{th}$ entry of the vector $\boldsymbol{\xi}^{(u)}$, we consider the median operator
\begin{equation*}
    \median\left(\boldsymbol{\xi}^{(u)}\right) = \begin{cases}
    \frac{\boldsymbol{\xi}^{(u)}_{[d_u/2]} +  \boldsymbol{\xi}^{(u)}_{[d_u/2 + 1]}}{2} & \text{for $d_u$ even}\\
    \boldsymbol{\xi}^{(u)}_{[(d_u+1)/2]} & \text{otherwise}.
    \end{cases}
\end{equation*}
\subsection{Fault Model}\label{sec:faultmodel}
We recall that $F$ is the set of malicious agents. Similarly to $V$, we define a partition of $F \subset V$:
\begin{equation*}
    F = \{F_1, F_2, \dots, F_c | F_i \subset \subs_i \ \text{and} \ |F_i| = f_i, \forall i \in [1, c]\},
\end{equation*}
Where we use $f_i$ to denote the number of malicious agents in the specific subset $\subs_i$.
From the literature, we present three types of fault models: \emph{F-total}, \emph{F-local} and \emph{Byzantine} models. The first two models require that at most $|F|$ malicious agents be present in the whole network (F-total) or in an agent's neighborhood (F-local) at any point in time. In the latter case, agents are still organized under an F-total or F-local scheme, though they are Byzantine agents. Our malicious agents are Byzantine, in that they can coordinate to strategically send a specific set of values to their neighbors.
Our model is
\emph{F-total} with respect to the whole agent set $V$.
\subsection{Community Structure}
As mentioned above, we look at a  graph $G$ within which there can be multiple regions satisfying specific connectivity constraints. We call these regions \emph{communities} and formally introduce them in  \cref{def:community}.
\begin{definition}[$(\kedges_i, f_i)$-community] \label{def:community}
    Consider a graph $G = (V, E)$ and a subset of agents $\subs_i \subseteq V$. For $\kedges_i, f_i \ge 0$, the induced sub-graph $\ginduced{_i}$ is a $(\kedges_i, f_i)$-\emph{community} if:
    \begin{itemize}
        \item $\ginduced{_i}$ is $(\kedges_i, f_i+1)$-excess robust;
        \item Its minimum degree is $d_{min}^{\subs_i} \ge 2f_i + \kedges_i + 1$
    \end{itemize}
\end{definition}

\cref{def:community} characterises a sub-graph of $G$ that is $(r,s)$-excess robust, however it does not imply anything about the robustness of $G$.
Note that, if $G$ is $(r,s)$-excess robust, then $\subs_i = V$, as well as the fact that communities can contain smaller communities within themselves. Furthermore, if $\ginduced{_i}$ is a $(\kedges_i, f_i)$-community, for $\subs_i' \subset \subs_i$, not necessarily is $\ginduced{_i'}$ a $(\kedges_i, f_i)$-community. For example, take $f_i' = 0, r > 0$. There might exist a subset $\subs_i' = \{u, v\}$ such that $(u, v) \notin E$: in this case, $\ginduced{_i'}$ is clearly not $(r, 1)$-excess robust, however, $u$ and $v$ may be connected to other agents in $\subs_i$ so that robustness of $\ginduced{_i}$ may be guaranteed all the same. Conversely, for $\subs_1, \subs_2 \subset V$ such that $\subs_1 \cap \subs_2 = \emptyset$ and for $\ginduced{_1}$ and $\ginduced{_2}$ two communities, $\ginduced{_1 \cup \subs_2}$ is not necessarily a community. A trivial example is two $(\kedges_i, f_i + 1)$-excess robust graphs being disconnected from each other.
\section{The Median Consensus Algorithm and \texorpdfstring{\gls{rac}}{}}\label{sec:algorithm}
We apply the same algorithm presented in \cite{medianSundaram} to the graph $G$ as described in \cref{sec:graph}.
Every agent $u \in V$ starts with a value $\xi_u(0) \in \mathbb{R}$, such that,
at any subsequent timestep $t$, the value $\xi_u(t)$ follows the same update step presented in the \gls{mca} introduced in \cite{medianSundaram}. Namely, for $\alpha \in (0, 1) $
\begin{equation} \label{eq:updatestep}
    \xi_u(t+1) =
        \resizebox{.45\hsize}{!}{$\alpha\xi_u(t) + (1-\alpha)\median(\boldsymbol{\xi}^{(u)}(t))$} \quad \text{if} \ u \in L
\end{equation}
This update step dictates that all the legitimate agents in the network will update their value to an average between their current value and the median of their neighbourhood. This is done synchronously for every agent $u \in L$. No assumptions are made on the update of the values for the malicious agents, nor on the value that they communicate to legitimate agents.

We define $\xi_{m_i}$ and $\xi_{M_i}$ as, respectively, $\min\{\xi_u(0)~|~u \in \subs_i\}$ and $\max\{\xi_u(0)~|~u \in \subs_i\}$. We therefore recall the definition of \emph{\gls{rac}} from \cite{medianSundaram} and expand it to account for this new notation.
\begin{definition}[Resilient Asymptotic Consensus]\label{def:resilientconsensus}
    Under any of the fault models and for a subset $\subs_i \in V$, the legitimate agents in $\subs_i$ are said to achieve \emph{\gls{rac}} if \emph{both} of the following conditions are satisfied for \emph{any} choice of initial values $\xi_u(0) \in \mathbb{R}$.
    \begin{itemize}
        \item \emph{Agreement Condition}: there exists $\fixedpoint{_i} \in \mathbb{R}$ such that $\lim_{t \to \infty}\xi_u(t) = \fixedpoint{_i}$;
        \item \emph{Safety Condition}: The values of agents are throughout between the minimum and maximum values of the legitimate agents, i.e., 
        $ for \ t \in \mathbb{Z}_{\ge 0}$, 
        $\xi_u(t) \in [\xi_{m_i}, \xi_{M_i}]$
    \end{itemize}
\end{definition}
\section{Theoretical Results} \label{sec:results}
The framework we introduced in the previous sections allows us to now present the results of our work .
We start by stating  the core result in \cref{thm:mainresult}, followed by an insight on the structure of its proof. \cref{thm:rsconstruction} follows, providing us with all the elements needed to later prove \cref{thm:mainresult}. 
For $u \in \subs_i$, recall that $\kedges_i = \max_u\{|\mathcal{N}_u \setminus \subs_i|\} \ge 0$ is the number of edges going from the node $u$ to a set outside of $\subs_i$. We call these ``external edges.''
\begin{theorem}\label{thm:mainresult}
    Take $G = (V, E)$, and the partition of $V$ in $c$ subsets $\subs_1, \subs_2, \dots, \subs_c$.
    For a given $\subs_i$ with $f_i$ malicious agents, and where each $u \in \subs_i$ has at most $\kedges_i \ge 0$ external edges,
    if $\ginduced{_i}$ is a $(\kedges_i, f_i)$-community, i.e.,
    \begin{enumerate}
    \item $\ginduced{_i}$ is $(\kedges_i, f_i + 1)$-excess robust
    \item $\ginduced{_i}$ has $d_{min}^{\subs_i} \ge 2f_i +  \kedges_i + 1$
    \end{enumerate}
    holds, and
    \begin{enumerate}
    \setcounter{enumi}{2}
    \item All legitimate agents in $\ginduced{_i}$ run \gls{mca}
    \end{enumerate}
    Then, every legitimate agent in $\subs_i$ will reach Resilient Asymptotic Consensus (cf.  \cref{def:resilientconsensus}).
\end{theorem}
\begin{remark}
Agents of different communities may converge to different values.
\end{remark}
\begin{remark}
    If $\kedges_i = 0$ and $c = 1$, which means that the graph consists of one single big community, this theorem is equivalent to Theorem 1 in \cite{medianSundaram}.
\end{remark}

\cref{thm:mainresult} shows that the use of \gls{mca} allows for the relaxation of two core prerequisites needed when applying it to reach community consensus. First, $G$ does not need to be $(r, s)$-excess robust in order to have subsets of agents reach consensus within $\ginduced{_i}$ through \gls{mca}; instead, it needs to have subsets of agents to be \emph{connected enough} by their induced sub-graphs being $(\kedges_i, f_i+1)$-excess robust.
Second, the minimum degree requirement for a node $u \in \subs_i$ is only dependent on the number of malicious agents that are in $\subs_i$ and the number of neighbors $u$ has outside of its community. This is much less restrictive than the degree requirement for global consensus, which is dependent on the total number of malicious agents in the entire network.
This implies that communities with fewer agents can also have a lower minimum degree. Simulations in \cref{sec:simulations} further confirm the theoretical results and the necessity of our assumptions.

\subsection{Proof Idea}
The full proof of \cref{thm:mainresult} can be found in \cref{sec:thmproof} and is structured as follows. We consider a graph $G$, as formalized in \cref{sec:graph}, and an induced sub-graph $\ginduced{_i}$. We show the following (\cref{thm:rsconstruction}). 
Consider  $\subs_i$: The agents within $\subs_i$ that are highly connected to other agents within the same subset still have more connections to agents within $\subs_i$ compared to outside $\subs_i$ if we look at $G$ instead of $\ginduced{_i}$.
More formally, if $\ginduced{_i}$ is a $(\kedges_i, f_i)$-community, then we show that agents that are $\kedges_i$-excess reachable in $\ginduced{_i}$ maintain high-enough reachability in $G$ -- even when connected to $\kedges_i$ new neighbors outside $\subs_i$. Therefore, we can apply the results from \cite{medianSundaram} (cf. Theorem 1 \cite{medianSundaram}) to show that agents in the community $\ginduced{_i}$ will reach resilient asymptotic consensus.


\subsection{Formal Analysis}
We start by defining the \emph{isolation} property of a community: it characterizes the case where agents in $\subs_i$ can be  connected to agents outside of $\subs_i$, i.e., $v, w, \ldots \in V \setminus \subs_i$, without resulting in the induced sub-graph being a community (cf. \cref{def:community}).
\begin{definition}[Community Isolation] \label{def:subIsolation}
Consider a community $\ginduced{_i}$, an agent $u\in \subs_i$ and a neighboring agent $v \in \mathcal{N}_u\setminus\subs_i$. $\ginduced{_i}$ is \emph{isolated} if, in the update step in \cref{eq:updatestep} for $u$, any $\xi_v(t)$ that does not satisfy the \emph{safety condition} does not influence the median operator.
\end{definition}
Making use of this, the following proposition characterizes how the excess reachability of agents within a community compares to the excess reachability within the graph. 
\begin{proposition}\label{thm:rsconstruction}
Take $G= (V,E)$ and a partition of $V$ into $c$ subsets $\{\subs_1, \subs_2, \dots, \subs_c\}$.
Let the induced sub-graph $\ginduced{_i}$ be a $(\kedges_i, f_i)$-community. Let $u$ be a $\kedges_i$-excess reachable agent in $\ginduced{_i}$ w.r.t. a set $S \subset \subs_i$. Then, $u$ is $0$-excess reachable in $G$ w.r.t. all sets $S \cup N$, where $N \subseteq \mathcal{N}_u \setminus \subs_i$. Furthermore, $\ginduced{_i}$ is isolated.
\end{proposition}
Before we prove this formally, it is worth emphasizing that the statement is non-trivial: First, agents in $\subs_i$ do not necessarily maintain the same reachability when looked at w.r.t. the whole graph $G$. Even in the absence of malicious agents, new edges connecting to external values could influence an agent $u \in \subs_i$ by making it diverge from the  $\fixedpoint{_i}$. Second, having more edges potentially means more malicious agents influencing a legitimate agent. It is therefore paramount that the cross-community edges do not expose any agent in a community to too many malicious agents.

\begin{figure}
    \centering
    \includegraphics[width=.5\textwidth]{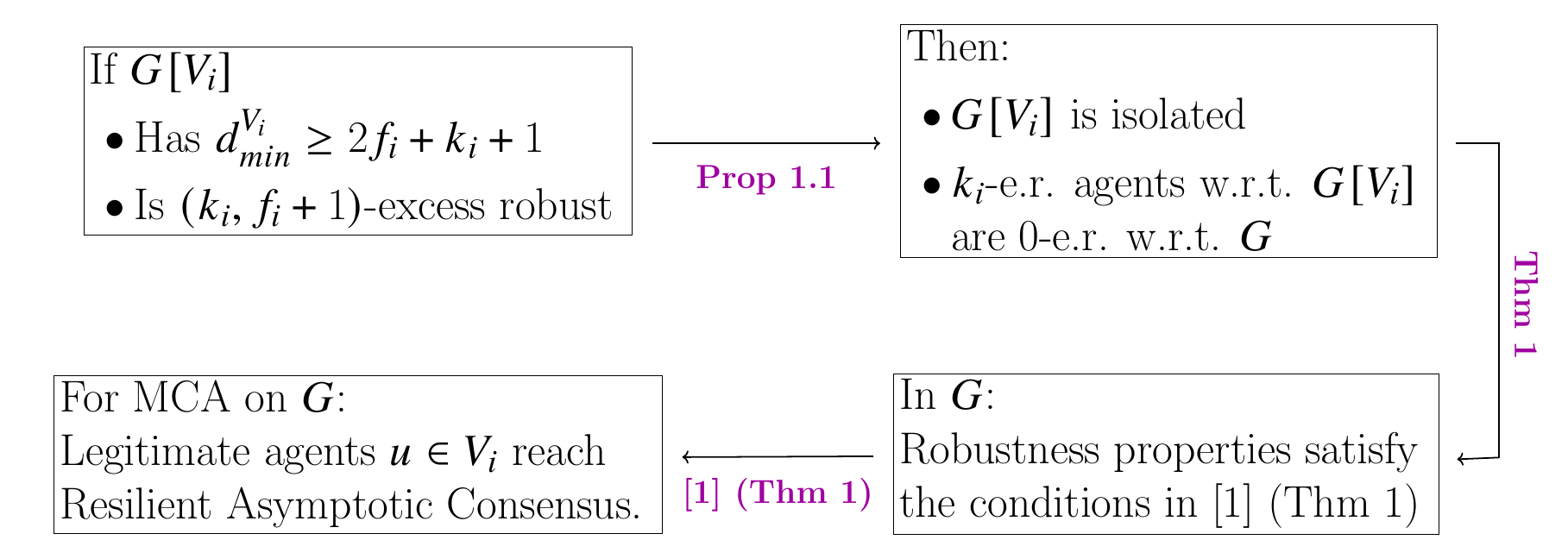}
    \caption{Flow of the proofs.}
    \label{fig:proofSketch}
    \vspace{-20pt}
\end{figure}
\begin{figure}
    \centering
    \begin{minipage}{0.47\columnwidth}
        \includegraphics[width=.77\textwidth]{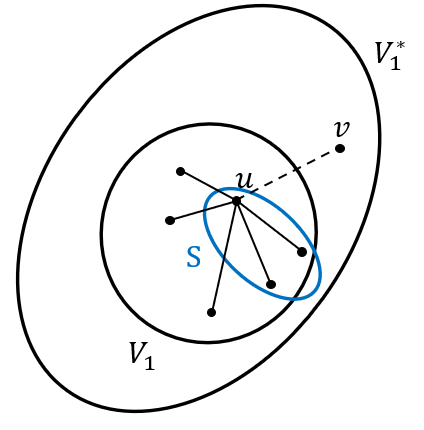}
    \end{minipage}
    \begin{minipage}{0.47\columnwidth}
        \includegraphics[width=.8\textwidth]{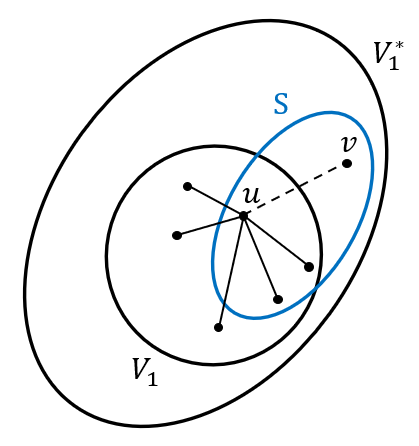}
    \end{minipage}
    \caption{Example of \emph{Case1} (left) and \emph{Case2} (right) in \cref{thm:rsconstruction}. With respect to $S$ and selecting the subset $\subs_i$, agent $u$ is $3-2 = 1$-excess reachable. When considering $\subs_i^*$, one agent more is outside $S$ and $u$ becomes $2$-excess reachable. In \emph{Case2}, $u$ becomes $3-3 = 0$-excess reachable.}
    \label{fig:case2b}
    \vspace{-10pt}
\end{figure}

\begin{proof}[Proof of \cref{thm:rsconstruction}] 
In the first part we prove that the agents that are $0$-excess reachable w.r.t. $\ginduced{_i}$ are $0$-excess reachable w.r.t. $G$.
In the second part, we show that $\ginduced{_i}$ is isolated.
\paragraph{Robustness Property}
Without loss of generality, we exclude the trivial case for which $\subs_i = V$ and $\kedges_i = 0$, and consider the community $\ginduced{_i}, \ \subs_i \subset V$. We take an agent $u \in \subs_i$ and an agent $v \in V \setminus \subs_i$,
and observe the connectivity of the induced sub-graph $\ginduced{_i^*} = \ginduced{_i\cup \{v\}}$.
By definition, any $r$-excess reachable agent is also $0$-excess reachable.
Key in this part is to show that any change in the connectivity of $\ginduced{_i^*}$ will not lead to the number of $0$-excess reachable agents in $\subs_i$ to decrease.
We therefore take a subset of agents $S \subseteq \subs_i^*$, such that $u \in S$ and that $S \cap \subs_i \neq \emptyset$. Without loss of generality, we assume that $u$ is a $\kedges_i$-excess reachable agent in $\ginduced{_i}$, given the choice $S$, and that $u$ and $v$ are connected by an edge. We observe that there are $2$ cases of interest:
1) $v \notin S$
   2) $v \in S$.
In the first case, $v$ being external to $S$ increases the count of external neighbors, leading to $\kedges_i + 1 > \kedges_i$ and maintaining $0$ reachability.
In case 2) both $u$ and $v$ belong to $S$, therefore the count of internal neighbors is increased by $1$. This yields $\kedges_i - 1 \ge 0$, since the exclusion of the case $\subs_i = V$ implies $\kedges_i \ge 1$.
Extending to the case where $\kedges_i$ agents external to $\subs_i$ are connected to $u$, we observe that, at worst case, $S$ contains all $\kedges_i$ external agents, yielding $\kedges_i - \kedges_i \ge 0$.
Any set of $\kedges_i$ additions is independent of each other, thus not affecting the reachability of the agents in $\subs_i$ with respect to each other. Additionally, for any agent $v$ belonging to a different community $\ginduced{_{j \neq i}}$, the same reasoning applies from $v$'s perspective, in this way proving the \emph{robustness property}.
See \cref{fig:case2b} for a graphic representation the two cases.
\paragraph{Isolation property}
Consider again the community $\ginduced{_i}$ and $\kedges_i$ external agents connected to $u \in \subs_i$. For $\ginduced{_i}$ to be isolated, we need to guarantee that the value $\xi_u(t) \in [\xi_{m_i}, \xi_{M_i}]$, respecting the \emph{safety condition}. 
It is sufficient to show that the median of the values from $u$'s neighbors is $M(\boldsymbol{\xi}^{(u)}(t)) \in [\xi_{m_i}, \xi_{M_i}]$. 
We prove this in the following.
By hypothesis, we know that $d_u^{\subs_i} \ge d_{min}^{\subs_i} \ge 2f_i + \kedges_i + 1$. By our model assumptions on the structure of $G$, any agent $u \in \subs_i$ can be connected to at most $\kedges$ agents that are external to $\subs_i$. Thus, the degree of $u$ with respect to the whole $G$ is $d_u = d_u^{\subs_i} + \kedges_i$ yielding
\begin{equation}\label{eq:degree}
\begin{split}
    d_u &= d_u^{\subs_i} + \kedges_i 
    \ge d_{min}^{\subs_i} + \kedges_i\\
    &\ge 2f_i + 2\kedges_i + 1 \ge 2(f_i+\kedges_i) + 1.
\end{split}
\end{equation}
The lower bound on $d_u$ in \cref{eq:degree} guarantees that the median operator will not be affected by any value $\xi_v(t) \notin [\xi_{m_i}, \xi_{M_i}]$. This includes both values from legitimate agents that are outside the interval $[\xi_{m_i}, \xi_{M_i}]$ and any value from external malicious agents. This proves isolation and the proposition.
\end{proof}

\subsection{Proof of \texorpdfstring{\cref{thm:mainresult}}{Theorem \cref{thm:mainresult}}}\label{sec:thmproof}

    Take a subset $\subs_i \subseteq V$ for which $\ginduced{_i}$ is a $(\kedges_i, f_i)$-community with $d_{min}^{\subs_i} \ge 2f_i + \kedges_i +1$.
    From \cref{thm:rsconstruction} we know that $\ginduced{_i}$ is isolated and that, the $f_i+1$ $\kedges_i$-excess reachable agents in $\ginduced{_i}$ will all be at least $0$-excess reachable in $G$. This means that all of these agents still connect to at least as many agents outside a chosen subset $S, \ s.t. \ S\cap \subs_i \neq \emptyset$, as agents inside it. Isolation guarantees that no outliers from agents outside of $\subs_i$ are selected. This satisfies the reachability conditions required from Theorem 1 in \cite{medianSundaram}, with respect to subset $\subs_i$. Precisely, that result leverages on the fact that $(0, f_i+1)$-robustness of the graph implies that there is always a combination of subsets $S_1, S_2$ for which enough agents are at least $0$-excess reachable. This concludes that \gls{mca} on $G$ will lead agents in $\subs_i$ to reach resilient asymptotic consensus and concludes the proof.
    $\qed$

\emph{Implications --}
The implications of \cref{thm:mainresult} and \cref{thm:rsconstruction} is twofold. On one hand, these results show that, even if consensus via \gls{mca} is not achieved globally, sub-graphs in the network which are connected enough can still agree on a common value. On the other hand, they give sufficient conditions to have a network of $(r, s)$-excess robust sub-graphs whose agreement to a value will not be hindered by being connected to each other. This is invaluable in many cases where several networks want to communicate with each other, yet still retaining the local information they converged to within themselves, and without compromising it, nor having it corrupted by further malicious agents.
Observe that these results do not imply anything about agents that do not belong to a community. In that case consensus may or may not be reached.

\section{Simulation Results} \label{sec:simulations}
\newcommand{\example}{Example}
In this section we complement our theoretical findings with simulations, and highlight cases where our conditions on the connectivity and degree are not met, resulting in community consensus failing. We show the successful case of community consensus when both the robustness and degree conditions are met, and two failure scenarios where one of these conditions is violated, respectively.
\subsection{Setup}
We start by describing the initial graph  $G$ that we then slightly modify to get the three cases.  
To obtain $G$, we start from two complete graphs $G_1=(V_1, E_1)$, $G_2=(V_2, E_2)$ with sizes $n_1$ and $n_2$, ($n_1 \ge n_2$ w.l.o.g.), and we set $\kedges_1 = \kedges_2 = \kedges$. 
Furthermore, we sample each $\xi_u(0)$ such that
\[
    \xi_u(0) \sim \begin{cases}
        N(2, 1) & \text{if} \quad u \in V_1 \cap L \\
        N(30, 5) & \text{if} \quad u \in V_2 \cap L \\
        60 & \text{if} \quad u \in F,
    \end{cases}
\]
where $N(\mu, \sigma^2)$ is the normal distribution with mean $\mu$ and variance $\sigma^2$.
As an example of an effective disruptive behaviour, we assign the same value to all the malicious agents and keep that value constant.
We consider three different setups:
\begin{itemize}
    \item $\ginduced{_1}$ and $\ginduced{_2}$ are $(\kedges, f_i)$-communities, this is our nominal case;
    \item At least one of the subgraphs $\ginduced{_i}$ does not respect robustness constraints, that is, it is not $(\kedges_i, f_i + 1)$-excess robust;
    \item At least one of the subgraphs $\ginduced{_i}$ does not respect the minimum degree condition $d_{min}^{\subs_i} \ge 2f_i +  \kedges_i + 1$.
\end{itemize}

\paragraph{\example 1: Constraints Respected}
We take $f_1 = 20$, $f_2 = 10$, and $n_1 = 6f_1 + 3$, $n_2 = 3f_2 + 5$. We set $\kedges = 2$. It can be verified that each complete graph is $(2, f_i)$-excess robust and the graph $G$ obtained in this way respects the constraints in \cref{thm:mainresult}. 
Note that the choice of $n_i$, as well as the parameters of the distributions and the value for the malicious agents, is completely arbitrary, as long as the minimum degree condition $d_{min} = d \ge 2f_i + 2 + 1$ is respected.

\begin{figure}[ht]
    \centering
    \includegraphics[width=.4\columnwidth]{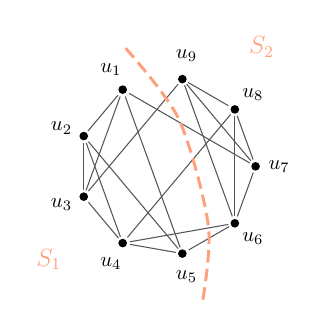}
    \caption[The Induced Subgraph used for Simulations in \cref{sec:theMedianConsensusProject}]{Graph used to simulate the case where robustness constraints are not met. The dashed line separates $\subs_2$ in two subsets of agent: $S_1 = \{u_1, \dots, u_5\}$ and $S_2 = \{u_6, \dots, u_9\}$. Recall that, in this case, it is $f_2 = 1$ and $\kedges_2 = 1$. Therefore it has to be $d_{min}^{\subs_2} \ge 2f_2 + \kedges_2 + 1 = 4$. In this example, no agent in $\subs_2$ is $1$-excess reachable under the partition $\{S_1, S_2\}$, thus $\ginduced{_2}$ is not $(1, 2)$-excess robust.}
    \label{fig:graphExample}
\end{figure}

\paragraph{\example 2: Robustness not Respected}
We set here $f_1 = 6$, $f_2 = 1$, $n_1 = 2f_1 + 4 = 16$ and $n_2 = 4f_2 + 5 = 9$. In this case we set $\kedges = 1$. By removing a number of edges from the $9$-clique, we obtain the sub-graph $\ginduced{_2}$ shown in \cref{fig:graphExample}. This network is not $(1, 2)$-excess robust, and one selection of subsets $S_1$ and $S_2$ that breaks the conditions for robustness is the subsets $S_1 = \{u_1, u_2, u_3, u_4, u_5\}$ and $S_2 = \{ u_6, u_7, u_8, u_9\}$, highlighted in \cref{fig:graphExample} by the dashed line. Nevertheless, the minimum degree condition is respected, since every agent in $\ginduced{_2}$ has at least degree $4$.


\paragraph{\example 3: Min Degree not Respected}
We set $f_1 = 6$, $f_2 = 3$, $n_1 = 2f_1 + 3 = 15$ and $n_2 = 2f_2 + 5 = 11$ agents each, and set $\kedges = 2$. The minimum degree condition for $\ginduced{_1}$ would be $d_{min}^{\subs_1} \ge 2f_1 + \kedges + 1 = 15$, which however cannot be respected since every node in $\ginduced{_1}$ has degree $d = 14$. On the other hand, the reader can verify that $\ginduced{_1}$ is $(2, 7)-$excess robust.

\paragraph{Code} \noindent The code for our experiments (\emph{Phase1.py}) can be found here: \url{https://bitbucket.org/CrissGava/majorityconsensuscode/src/master/}
\subsection{Findings}
\begin{figure*}[t!]
    \centering
    \begin{subfigure}{.32\textwidth}
    \includegraphics[width=\textwidth]{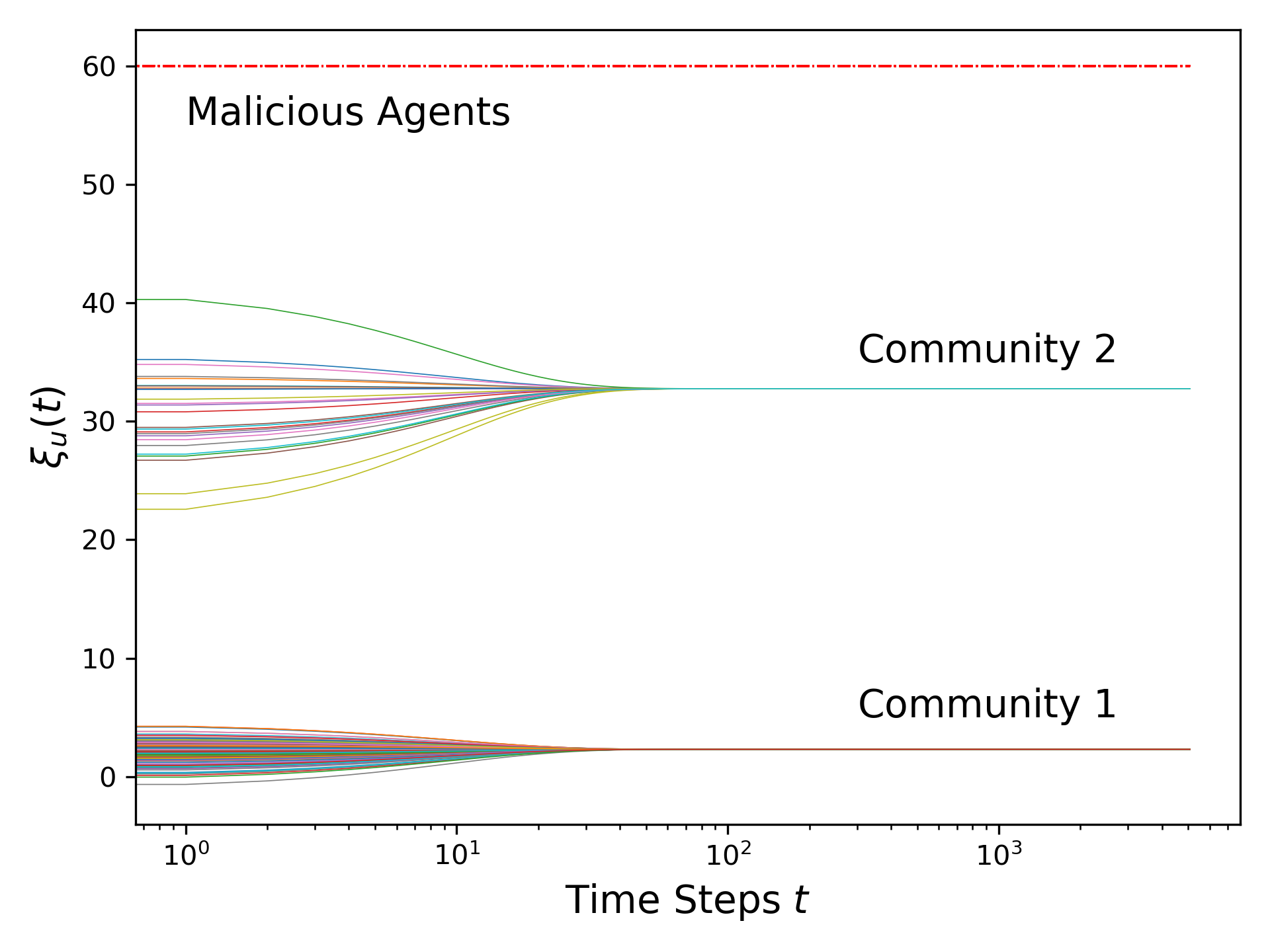}
    \caption{Community Consensus reached}
    \label{fig:allGood}
    \end{subfigure}
    \begin{subfigure}{.32\textwidth}
    \includegraphics[width=\textwidth]{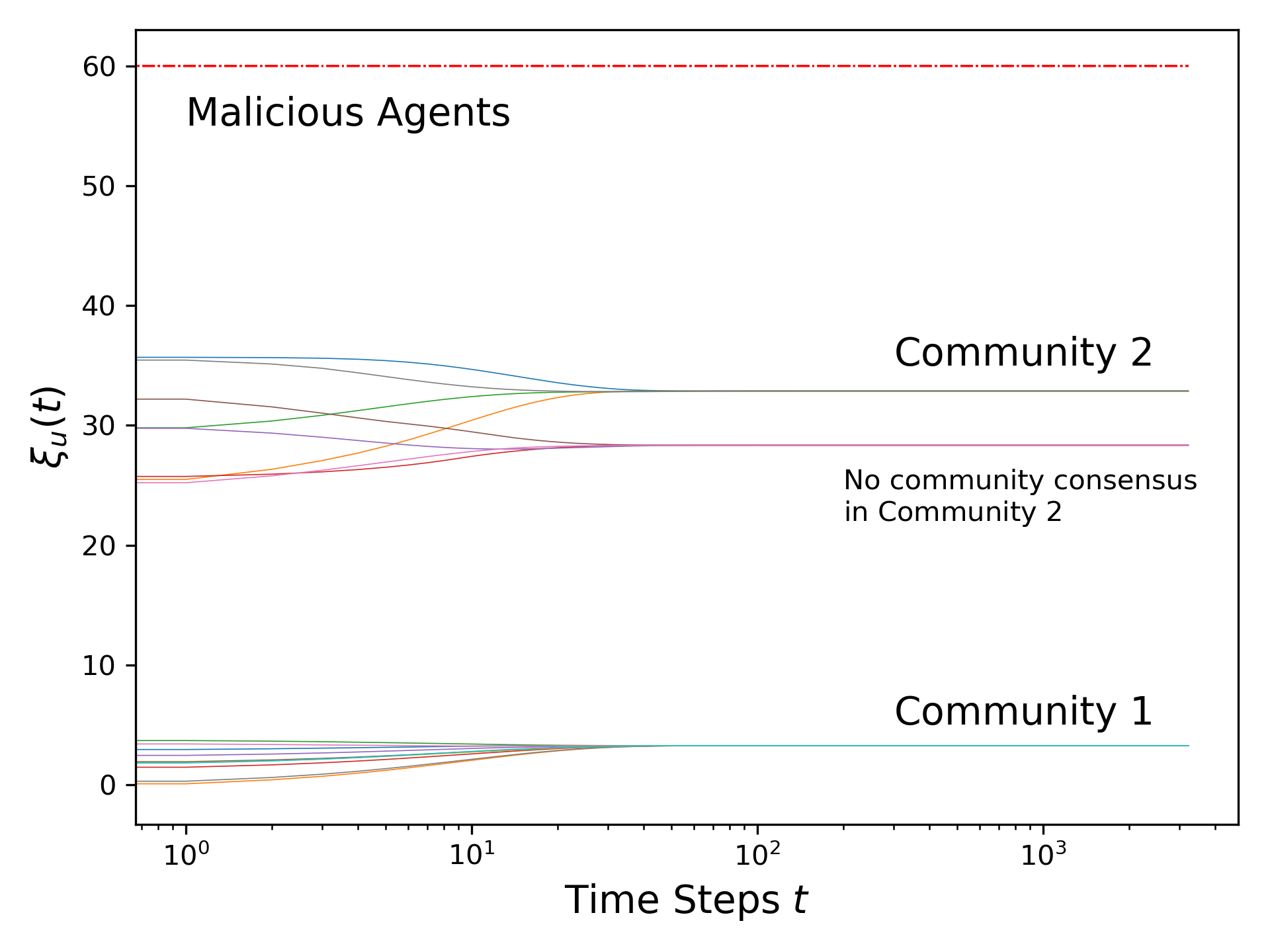}
    \caption{Robustness constrain violated}
    \label{fig:noRobustness}
    \end{subfigure}
    \begin{subfigure}{.32\textwidth}
    \includegraphics[width=\textwidth]{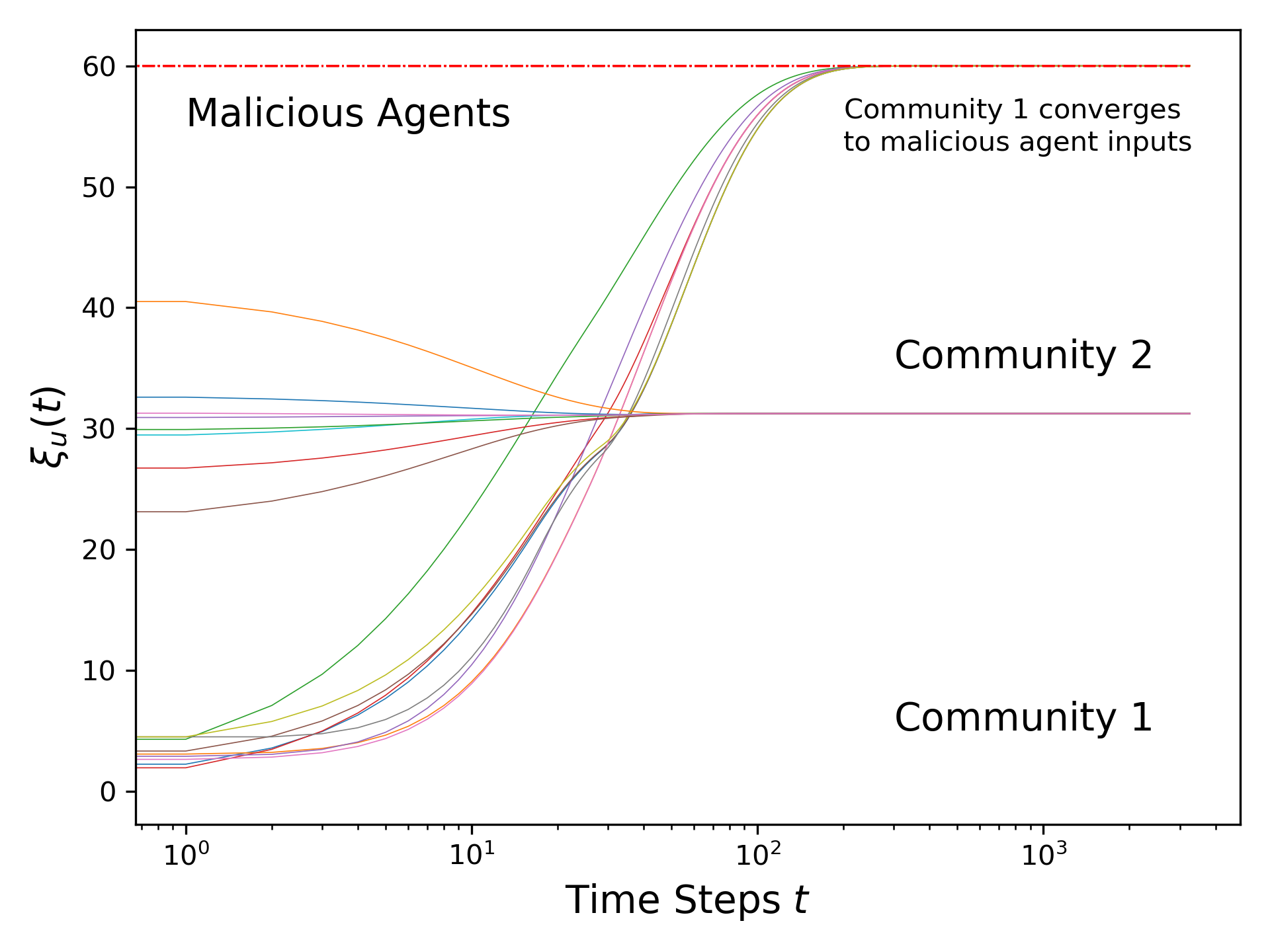}
    \caption{Min degree constraint violated}
    \label{fig:noDegree}
    \end{subfigure}
    \caption{Plots of our simulation results (cf. \cref{sec:simulations}). (a) - \emph{\example 1}: Constraints from \cref{thm:mainresult} are respected, (b) - \emph{\example 2}: Robustness constraint is violated: agents in $\ginduced{_2}$ converge to two different values, (c) - \emph{\example 3}: Minimum degree constraint is violated. Only in (a) do legitimate agents reach community consensus. The value of 60 belongs to malicious agents and that the x-axis uses logarithmic scale.}
\end{figure*}

The three plots in \cref{fig:allGood}--\cref{fig:noDegree} show the outcomes of each example. Each line in the plot represents the value of an agent. Note, that the convergence rate is intentionally set to be very slow by setting $\alpha = 0.9$ in the consensus update rule. This is done to provide more insight into the convergence process.

{\bf Successful community consensus --} When the conditions of \cref{thm:mainresult} are met, community consensus is reached, as shown in \cref{fig:allGood}.
At $t=0$ all the agents start with different values, and they quickly converge within their communities. 

{\bf Failure due to violation of robustness --} 
Not respecting the robustness condition means that if the malicious agents are positioned strategically in important nodes of the graph, they can prevent information flow between some subsets of the legitimate agents. Consequently, those subsets -- depending on their initial values -- might converge to different values from each other. This is exactly what happens in $\ginduced{_2}$ in \cref{fig:noRobustness}. Notice, that since the degree condition is respected, all of the values of the legitimate agents stay within the convex hull of their initial values.

{\bf Failure due to violation of minimum degree --} 
When the minimum degree condition is not respected in $\ginduced{_i}$, legitimate agents can include values of malicious agents, or legitimate agents from outside $\ginduced{_i}$ in their update rule. Whether this happens or not is dependent on both the initialization of all agents and the placement of the malicious agents. The effects of such inclusion can result in legitimate agents in $\ginduced{_i}$ possibly converging to values outside the convex hull of their initial values. This is showcased in \cref{fig:noDegree}, with agents in $\ginduced{_1}$ converging to the values of the malicious agents.


\section{Practical Use Cases} \label{sec:usecases}
We argue that Community Consensus can be applied to several real-life use cases. Heterogeneous connectivity in networks is a common occurrence, and many theoretical models account for it, from scale-free, or assortative networks, to different agent centrality measures and clustering criteria. We now present here two use cases

{\it Distributed Sensing --} Multiple sensor networks are implemented for  various applications: from autonomous vehicle coordination, to energy infrastructure, or seismic and geothermal monitoring. These have steadily evolved to large swarms of micro-sensors and \gls{iot} \cite{abbasian2020survey}, creating new challenges in the handling of the big amount of data subject to infrastructure constraints, such as bandwidth limitations and processing power. Different solutions are being explored, from mobile-agent-based \gls{dsn} and sensor fusion algorithms \cite{dsnQI2001655}, to \gls{das} to map ocean floors \cite{acoustic2019}. 
Because of the extensive surface that a \gls{das} application to environmental sensing covers and the extreme diversity of the maritime ecosystem, one might want to segregate the data related to different regions of the ocean floor.

{\it Blockchain Sharding --} The blockchain technology is extremely powerful when it comes to approving transactions in a decentralized way and while guaranteeing security. This concept is being used in many applications, from finance to large scale \gls{iot} \cite{shardingCai}. Nonetheless, it is still computationally expensive and poorly scalable. A solution comes with \emph{sharding}, an approach to allocate different parts of a blockchain transaction to different \emph{shards}, or communities \cite{LIU2022100513, wang2019sok}. Shards are connected to each other, so to communicate their result and therefore complete the entire transaction. However, issues like shard invalidation can be caused by malicious agents in an even smaller number than what the entire blockchain would be robust to. Few works focused on optimizing related consensus algorithms yet, and even fewer focused on detailing the robustness of the model given by the network structure.
We believe that there is room for exploration and application of our work in these scenarios, reinforcing its versatility and power.
\section{Conclusions}\label{sec:conclusions}

In this work we presented a novel distributed framework, called community consensus, and gave conditions under which the \gls{mca} allows agents to reach consensus within their communities, even if some malicious entities are present. 

We show the importance of our conditions by providing settings which do not respect said conditions resulting in the agents failing to reach community consensus.  
%
This approach finds a place in practical applications and realistic scenarios, where a lower or heterogeneous connectivity is expected, or where  it is in fact necessary to maintain regions of a network connected, and at the same time have them able to converge independently of each other.


\newpage
\printbibliography

@ARTICLE{shardingCai,
  author={Cai, Xingjuan and Geng, Shaojin and Zhang, Jingbo and Wu, Di and Cui, Zhihua and Zhang, Wensheng and Chen, Jinjun},
  journal={IEEE Transactions on Industrial Informatics}, 
  title={A Sharding Scheme-Based Many-Objective Optimization Algorithm for Enhancing Security in Blockchain-Enabled Industrial Internet of Things}, 
  year={2021},
  volume={17},
  number={11},
  pages={7650-7658},
  doi={}
}

@inproceedings{wang2019sok,
  title={Sok: Sharding on blockchain},
  author={Wang, Gang and Shi, Zhijie Jerry and Nixon, Mark and Han, Song},
  booktitle={Proceedings of the 1st ACM Conference on Advances in Financial Technologies},
  pages={41--61},
  year={2019}
}

@article{LIU2022100513,
title = {Building blocks of sharding blockchain systems: Concepts, approaches, and open problems},
journal = {Computer Science Review},
volume = {46},
pages = {100513},
year = {2022},
doi = {},
url = {},
author = {Yizhong Liu and Jianwei Liu and Marcos Antonio {Vaz Salles} and Zongyang Zhang and Tong Li and Bin Hu and Fritz Henglein and Rongxing Lu}
}

@article{abbasian2020survey,
  title={A survey on data aggregation techniques in IoT sensor networks},
  author={Abbasian Dehkordi, Soroush and Farajzadeh, Kamran and Rezazadeh, Javad and Farahbakhsh, Reza and Sandrasegaran, Kumbesan and Abbasian Dehkordi, Masih},
  journal={Wireless Networks},
  volume={26},
  pages={1243--1263},
  year={2020},
  publisher={Springer}
}

@article{acoustic2019,
author = {Sladen, Anthony and Rivet, Diane and Ampuero, Jean Paul and Barros, Louis and Hello, Yann and Calbris, Gaëtan and Lamare, P.},
year = {2019},
month = {12},
pages = {5777},
title = {Distributed sensing of earthquakes and ocean-solid Earth interactions on seafloor telecom cables},
volume = {10},
journal = {Nature Communications},
doi = {}
}

@article{dsnQI2001655,
title = {Distributed sensor networks--a review of recent research},
journal = {Journal of the Franklin Institute},
volume = {338},
number = {6},
pages = {655-668},
year = {2001},
note = {Distributed Sensor Networks for Real-time Systems with Adaptive C onfiguration},
doi = {}}

@misc{berenbrink2023distributed,
      title={Distributed Averaging in Population Protocols}, 
      author={Petra Berenbrink and Colin Cooper and Cristina Gava and David Kohan Marzagão and Frederik Mallmann-Trenn and Nicolás Rivera and Tomasz Radzik},
      year={2023},
      eprint={2211.17125},
      archivePrefix={arXiv}
}

@INPROCEEDINGS{wmsrsundaramc,
  author={Zhang, Haotian and Sundaram, Shreyas},
  booktitle={2012 American Control Conference (ACC)}, 
  title={Robustness of information diffusion algorithms to locally bounded adversaries}, 
  year={2012},
  volume={},
  number={},
  pages={5855-5861},
  %doi={10.1109/ACC.2012.6315661}
  }

@INPROCEEDINGS{haotian2012,
  author={Zhang, Haotian and Sundaram, Shreyas},
  booktitle={2012 IEEE 51st IEEE Conference on Decision and Control (CDC)}, 
  title={Robustness of complex networks with implications for consensus and contagion}, 
  year={2012},
  volume={},
  number={},
  pages={3426-3432},
  %doi={10.1109/CDC.2012.6425841}
  }

@article{angluin2008simple,
  title={A simple population protocol for fast robust approximate majority},
  author={Angluin, Dana and Aspnes, James and Eisenstat, David},
  journal={Distributed Computing},
  volume={21},
  number={2},
  pages={87--102},
  year={2008},
  publisher={Springer}
}

@article{LORENZ2005217,
title = {A stabilization theorem for dynamics of continuous opinions},
journal = {Physica A: Statistical Mechanics and its Applications},
volume = {355},
number = {1},
pages = {217-223},
year = {2005},
publisher = {Market Dynamics and Quantitative Economics},
%issn = {0378-4371},
%doi = {https://doi.org/10.1016/j.physa.2005.02.086},
%url = {https://www.sciencedirect.com/science/article/pii/S0378437105002955},
author = {Jan Lorenz},
keywords = {Continuous opinion dynamics, Non-negative matrices, Repeated averaging, Positive diagonal},
abstract = {A stabilization theorem for processes of opinion dynamics is presented. The theorem is applicable to a wide class of models of continuous opinion dynamics based on averaging (like the models of Hegselmannâ€“Krause and Weisbuchâ€“Deffuant). The analysis detects self-confidence as a driving force of stabilization.}
}

@phdthesis{lorenz2007repeated,
  title={Repeated averaging and bounded confidence modeling, analysis and simulation of continuous opinion dynamics},
  author={Lorenz, Jan},
  year={2007},
  school={Universit{\"a}t Bremen}
}

@article{Degroot1974,
author = { Morris H.   Degroot },
title = {Reaching a Consensus},
journal = {Journal of the American Statistical Association},
volume = {69},
number = {345},
pages = {118-121},
year  = {1974},
publisher = {Taylor & Francis},
%doi = {10.1080/01621459.1974.10480137},

%URL = { https://www.tandfonline.com/doi/abs/10.1080/01621459.1974.10480137},
%eprint = { https://www.tandfonline.com/doi/pdf/10.1080/01621459.1974.10480137}
}

@ARTICLE{Jadbabaie2003,
  author={Jadbabaie, A. and Jie Lin and Morse, A.S.},
  journal={IEEE Transactions on Automatic Control}, 
  title={Coordination of groups of mobile autonomous agents using nearest neighbor rules}, 
  year={2003},
  volume={48},
  number={6},
  pages={988-1001},
  %doi={10.1109/TAC.2003.812781}
  }

@ARTICLE{OSaber2004,
  author={Olfati-Saber, R. and Murray, R.M.},
  journal={IEEE Transactions on Automatic Control}, 
  title={Consensus problems in networks of agents with switching topology and time-delays}, 
  year={2004},
  volume={49},
  number={9},
  pages={1520-1533},
  %doi={10.1109/TAC.2004.834113}
  }

@article{parlangeli2021detection,
  title={On the detection and identification of edge disconnections in a multi-agent consensus network},
  author={Parlangeli, Gianfranco and Valcher, Maria Elena},
  journal={arXiv preprint arXiv:2101.06728},
  year={2021}
}

@article{FMS22,
  author       = {Frederik Mallmann{-}Trenn and
                  Matthew Cavorsi and
                  Stephanie Gil},
  title        = {Crowd Vetting: Rejecting Adversaries via Collaboration With Application
                  to Multirobot Flocking},
  journal      = {{IEEE} Trans. Robotics},
  volume       = {38},
  number       = {1},
  pages        = {5--24},
  year         = {2022},
  doi          = {10.1109/TRO.2021.3089033},
  bibsource    = {dblp computer science bibliography, https://dblp.org}
}

@ARTICLE{Wang2019,
  author={Wang, Bohui and Chen, Weisheng and Wang, Jingcheng and Zhang, Bin and Zhang, Zhengqiang and Qiu, Xingguo},
  journal={IEEE Transactions on Cybernetics}, 
  title={Cooperative Tracking Control of Multiagent Systems: A Heterogeneous Coupling Network and Intermittent Communication Framework}, 
  year={2019},
  volume={49},
  number={12},
  pages={4308-4320},
  %doi={10.1109/TCYB.2018.2859345}
  }

@InProceedings{Angluin2006,
author={Angluin, Dana and Fischer, Michael J. and Jiang, Hong},
editor={Gibbons, Phillip B.
and Abdelzaher, Tarek
and Aspnes, James
and Rao, Ramesh},
title={Stabilizing Consensus in Mobile Networks},
booktitle={Distributed Computing in Sensor Systems},
year={2006},
publisher={Springer Berlin Heidelberg},
pages={37--50}
}

@INPROCEEDINGS{Cao2005,
  author={Cao, M. and Spielman, D.A. and Morse, A.S.},
  booktitle={Proceedings of the 44th IEEE Conference on Decision and Control}, 
  title={A Lower Bound on Convergence of a Distributed Network Consensus Algorithm}, 
  year={2005},
  volume={},
  number={},
  pages={2356-2361},
  %doi={10.1109/CDC.2005.1582514}
  }

@article{cao2008reaching,
  title={Reaching a consensus in a dynamically changing environment: A graphical approach},
  author={Cao, Ming and Morse, A Stephen and Anderson, Brian DO},
  journal={SIAM Journal on Control and Optimization},
  volume={47},
  number={2},
  pages={575--600},
  year={2008},
  publisher={SIAM}
}

@article{Cao2008,
author = {Cao, Ming and Morse, A. and Anderson, Brian},
year = {2008},
month = {01},
pages = {601-623},
title = {Reaching a Consensus in a Dynamically Changing Environment: Convergence Rates, Measurement Delays, and Asynchronous Events},
volume = {47},
journal = {SIAM J. Control and Optimization},
%doi = {10.1137/060657029}
}

@ARTICLE{Ming2008,
  author={Cao, Ming and Morse, A. Stephen and Anderson, Brian D. O.},
  journal={IEEE Transactions on Automatic Control}, 
  title={Agreeing Asynchronously}, 
  year={2008},
  volume={53},
  number={8},
  pages={1826-1838},
  %doi={10.1109/TAC.2008.929387}
}

@INPROCEEDINGS{Kempe2003,
author={Kempe, D. and Dobra, A. and Gehrke, J.},
booktitle={44th Annual IEEE Symposium on Foundations of Computer Science, 2003. Proceedings.},
title={Gossip-based computation of aggregate information},   year={2003},
volume={}, 
number={}, 
pages={482-491},
%doi={10.1109/SFCS.2003.1238221}
}

@inproceedings{Kuhn2007,
author = {Kuhn, Fabian and Locher, Thomas and Wattenhofer, Rogert},
title = {Tight Bounds for Distributed Selection},
year = {2007},
}

@inproceedings{Doerr2011,
author = {Doerr, Benjamin and Goldberg, Leslie Ann and Minder, Lorenz and Sauerwald, Thomas and Scheideler, Christian},
title = {Stabilizing Consensus with the Power of Two Choices},
year = {2011},
}

@article{vicsek1995novel,
  title={Novel type of phase transition in a system of self-driven particles},
  author={Vicsek, Tam{\'a}s and Czir{\'o}k, Andr{\'a}s and Ben-Jacob, Eshel and Cohen, Inon and Shochet, Ofer},
  journal={Physical review letters},
  volume={75},
  number={6},
  pages={1226},
  year={1995},
  publisher={APS}
}

@INPROCEEDINGS{medianSundaram,
  author={Zhang, Haotian and Sundaram, Shreyas},
  booktitle={2012 50th Annual Allerton Conference on Communication, Control, and Computing}, 
  title={A simple median-based resilient consensus algorithm}, 
  year={2012},
  volume={},
  number={},
  pages={1734-1741},
  %doi={10.1109/Allerton.2012.6483431}
  }

@article{blanchard2017machine,
  title={Machine learning with adversaries: Byzantine tolerant gradient descent},
  author={Blanchard, Peva and El Mhamdi, El Mahdi and Guerraoui, Rachid and Stainer, Julien},
  journal={Advances in Neural Information Processing Systems},
  volume={30},
  year={2017}
}

@article{yang2019byrdie,
  title={Byrdie: Byzantine-resilient distributed coordinate descent for decentralized learning},
  author={Yang, Zhixiong and Bajwa, Waheed U},
  journal={IEEE Transactions on Signal and Information Processing over Networks},
  volume={5},
  number={4},
  pages={611--627},
  year={2019},
  publisher={IEEE}
}

@article{chen2017distributed,
  title={Distributed statistical machine learning in adversarial settings: Byzantine gradient descent},
  author={Chen, Yudong and Su, Lili and Xu, Jiaming},
  journal={Proceedings of the ACM on Measurement and Analysis of Computing Systems},
  volume={1},
  number={2},
  pages={1--25},
  year={2017},
  publisher={ACM New York, NY, USA}
}

@inproceedings{guerraoui2018hidden,
  title={The hidden vulnerability of distributed learning in byzantium},
  author={Guerraoui, Rachid and Rouault, S{\'e}bastien and others},
  booktitle={International Conference on Machine Learning},
  pages={3521--3530},
  year={2018},
  organization={PMLR}
}

@article{pasqualetti2011consensus,
  title={Consensus computation in unreliable networks: A system theoretic approach},
  author={Pasqualetti, Fabio and Bicchi, Antonio and Bullo, Francesco},
  journal={IEEE Transactions on Automatic Control},
  volume={57},
  number={1},
  pages={90--104},
  year={2011},
  publisher={IEEE}
}

@article{yemini2021characterizing,
  title={Characterizing trust and resilience in distributed consensus for cyberphysical systems},
  author={Yemini, Michal and Nedi{\'c}, Angelia and Goldsmith, Andrea J and Gil, Stephanie},
  journal={IEEE Transactions on Robotics},
  volume={38},
  number={1},
  pages={71--91},
  year={2021},
  publisher={IEEE}
}

@article{gil2018resilient,
  title={Resilient multi-agent consensus using wi-fi signals},
  author={Gil, Stephanie and Baykal, Cenk and Rus, Daniela},
  journal={IEEE control systems letters},
  volume={3},
  number={1},
  pages={126--131},
  year={2018},
  publisher={IEEE}
}
\end{document}